\documentstyle[12pt,aasms4]{article}

\newcommand{\ha}{H$\alpha$}
\newcommand{\hb}{H$\beta$}
\newcommand{\hi}{\protect\ion{H}{1}}
\newcommand{\hii}{\protect\ion{H}{2}}
\newcommand{\sii}{[\protect\ion{S}{2}]}
\newcommand{\Ni}{[\protect\ion{N}{1}]}
\newcommand{\cii}{\protect\ion{C}{2}}
\newcommand{\ciii}{\protect\ion{C}{3}}
\newcommand{\ciibr}{\protect\ion{C}{2}]}
\newcommand{\ciiibr}{\protect\ion{C}{3}]}
\newcommand{\nii}{[\protect\ion{N}{2}]}
\newcommand{\oi}{[\protect\ion{O}{1}]}
\newcommand{\oii}{[\protect\ion{O}{2}]}
\newcommand{\oiii}{[\protect\ion{O}{3}]}
\newcommand{\kms}{km~s$^{-1}$}
\newcommand{\cc}{cm$^{-3}$}
\newcommand{\emm}{pc~cm$^{-6}$}
\newcommand{\ebv}{$E(B-V)$}

\begin{document}

\title{Emission-Line Properties of the LMC Bubble N70}

\author{Brooke P. Skelton}
\affil{Astronomy Department, University of Washington, Seattle, Washington
98195\\ e-mail: skelton@astro.washington.edu}

\author{William H. Waller\altaffilmark{1,2}}\affil{Raytheon STX Corporation, NASA
Goddard Space Flight Center\\ Laboratory for Astronomy and Solar Physics, Code
681, Greenbelt, Maryland 20771}

\author{Richard F. Gelderman\altaffilmark{3}}\affil{National Research Council \& NASA Goddard
Space Flight Center\\ Laboratory for Astronomy and Solar Physics, Code
681, Greenbelt, Maryland 20771}

\author{Larry W. Brown and Bruce E. Woodgate}\affil{NASA Goddard Space Flight
Center\\ Laboratory for Astronomy and Solar Physics, Code 681, Greenbelt,
Maryland 20771}

\author{Adeline Caulet\altaffilmark{1}}\affil
{Astronomy Department, University of Illinois, Urbana, IL 61801}

\and

\author{Robert A. Schommer}\affil{National Optical Astronomy Observatories, Cerro Tololo Inter-American Observatory\\Casilla 603, La Serena, Chile}

\altaffiltext{1}{Visiting Astronomer, Cerro Tololo Inter-American Observatory. 
CTIO is operated by AURA, Inc.\ under contract to the National Science
Foundation.} 

\altaffiltext{2}{Current Address:  Harvard Smithsonian Center for Astrophysics,
60 Garden Street, Cambridge, Massachusetts  02138}

\altaffiltext{3}{Current Address:  Western Kentucky University, Department of Physics and Astronomy, Bowling Green, Kentucky  42101-3576}

\begin{abstract}
We present a spectrophotometric imaging study of the emission bubble N70 (DEM
301) in the Large Magellanic Cloud.  N70 is approximately 100 pc in
size with a nearly circular shell-like morphology.  The nebular
emission is powered by an uncertain combination of EUV photons, intense
winds, and supernova shock waves from the central population of
high-mass stars (the OB association LH 114).  We have obtained
narrow-band images (FWHM $\sim$6\AA) of N70 in the light of
\ha$\lambda$6563, \nii$\lambda$6584, \sii$\lambda\lambda$6717,6731, and
\oiii$\lambda$5007, along with the corresponding red and green
continua.  The resulting line fluxes and flux ratios are used to derive
ionization rates, nebular densities, volume filling fractions, and
excitation indices.  The photoionizing luminosity inferred from the
embedded stellar population is more than adequate to account for the
observed hydrogen ionization rate.

We compare the emission-line photometry with that derived from similar
imaging of the Orion nebula and with data collected from the literature
on other emission-line regions in the LMC.  Compared to the Orion
nebula, N70 shows much higher \sii/\ha\ intensity ratios which increase
smoothly with radius --- from $<$~0.3 near the center to $>$~1.0
towards the outer filamentary shell.  The measured intensity ratios in
N70 more closely match the range of excitation spanned by
 giant and supergiant \hii\ shells and by some of the supernova
remnants observed in the LMC.   The contending ionization and
excitation processes in the interior and outer shell of N70 are
evaluated in terms of the available data.  EUV photons probably
contribute most of the inner nebula's ionization, whereas a combination
of photoionization plus collisional ionization and excitation of sulfur
atoms by low-velocity shocks seems to best fit the emission-line
luminosities and intensity ratios observed in the outer shell.
Considerations of the radiative and mechanical energetics that are
involved may indicate the need for one or two supernova explosions
having occurred during the last $\sim$ Myr.

\end{abstract}


\section{Introduction}

N70 (DEM 301; 5$^{\mbox{h}}$43$^{\mbox{m}}$16$^{\mbox{s}}$
$-$67\arcdeg50\arcmin53\arcsec\ (J2000)) in the Large Magellanic Cloud is an
especially prominent bubble of line-emitting gas which appears to be powered
by a population of hot massive stars in its interior.  Its nearly circular
symmetry has prompted several studies aimed at determining which mechanisms
govern the dynamical and radiative evolution of this seemingly isolated and
``simple'' starburst.  An early spectrophotometric study by Dopita et al.\
(1981) led to the conclusion that N70 represents a giant cavity that has been
blown into a massive \hi\ cloud by the embedded cluster of hot, windy stars.
Ultraviolet radiation from the O stars was regarded as the dominant
source of ionization in the shell.

Lasker (1977) suggested that additional excitation by stellar winds is
necessary to explain the high \sii/\ha\ line ratios that he observed.
The detection of lopsided X-ray emission from the southwestern quadrant
of N70 indicated to Chu and Mac Low (1990) that a supernova explosion
occurred inside a wind-blown bubble.  The energy from such a supernova
would have provided additional heating of the gas inside the bubble,
accounting for the higher-than-expected X-ray emission, and could have
shocked the pre-existing shell of warm gas in the visible shell, raising its
\sii/\ha\ line ratio.

More recently, Oey (1996a, 1996b) has attempted to explain the dynamics
of N70 with standard pressure-driven bubble models; in doing so, she
has examined the stellar population of the central OB association, LH
114 (\cite{luc70}).  The current population of stars in LH 114 has an
initial mass function (IMF) slope consistent with a Salpeter IMF
slope.  If only the stars interior to the edge of the nebula are
considered, a slightly flatter slope of $\Gamma \sim -1.0$ is
determined ($\Gamma_{Sal}=-1.35$).  The mean age of these stars is
approximately 5 million years; three stars are approximately 40
M$_{\odot}$ and none are more massive (\cite{oey96a}).

Based on this inferred age, Oey found that N70 is a ``high-velocity''
superbubble which is dynamically inconsistent with the standard
model---the expansion velocities observed in N70 being too high for the
observed radius and age (\cite{oey96b}).  This, of course, assumes that
the motions observed in N70 are in fact due to expansion, which Dopita
et al.\ (1981) contests.  Kinematics of N70 have been observed by
Dopita et al.\ (1981), Rosado et al.\ (1981), Blades et al.\ (1980),
Georgelin et al.\ (1983), and Lasker (1977), yielding velocity
dispersions of order 40 \kms\ but differing conclusions regarding the
overall velocity field.

In this paper, we present narrow-band images of the ionized bubble in
the light of \ha, \sii, \nii, and \oiii.  These images were taken at
the CTIO 1.5-m telescope with the Rutgers Fabry Perot imaging system
and Goddard Fabry Perot etalons whose tunable narrow-band capability
enables the clean separation of the \ha$\lambda6563$ and
\nii$\lambda$6584 lines as well as the doublet lines of
\sii$\lambda$6717 and \sii$\lambda$6731.  The resulting line flux and
flux ratio measurements are used to derive nebular ionization rates,
densities, volume filling fractions, and indices of excitation.
Analysis of these spectrophotometric indices and comparison with
similar data on the Orion nebula and other line-emitting sources in the
LMC indicates that a {\it mix} of radiative and collisional (shock)
processes is responsible for ionizing and exciting the gas in this
shell.  Empirically, the N70 flux ratios span a range of values similar
to that of other LMC giant and supergiant shells except for a few
regions with unusually high \sii\ that are more characteristic of
supernova remnants.  A comparison with shock models
(e.g.\ \cite{shu79}, \cite{har94}) indicates that the high
\sii/\ha\ line ratios and the low, Orion-like \nii/\ha\ line ratios in
N70 are difficult to reconcile with a hybrid mode of nebular
photoionization and shock-excitation without invoking special
circumstances.

The Fabry-Perot observations of N70 and the data reduction are
described in Section \ref{obs}.  Images of the \ha, \nii, \sii, and
\oiii\ line emission and of their intensity ratios are presented and
discussed in Section \ref{maps}.  Comparisons with the Orion Nebula,
other line emitting regions in the LMC, and nebular models are made in
Section \ref{or_neb}, where constraints on the total radiative and
mechanical energetics are considered.  A summary of our results is
presented in Section \ref{summary}.

\section{Fabry-Perot Observations and Reductions \label{obs}}

Compared to the previous spectrophotometric studies of N70
(\cite{dop81}, \cite{las81}), the observations presented herein
represent an improvement in sensitivity and multi-spectral coverage at
\ha, \nii, \sii, and \oiii, including resolution of the \sii\ doublet,
thereby enabling a more detailed and comprehensive analysis of the
nebular ionization and excitation.  While previous studies have
presented images of N70 in the light of \ha, none have imaged the
entire nebula in \nii, \sii, or \oiii\ using CCD technology or the very
narrow bandwidths allowed by Fabry-Perot cameras.  While the
spectroscopy employed by these studies allows one to obtain kinematic
information, the present observations are the first to afford a
comprehensive look at how the emission line fluxes and ratios change on
small scales ($< 1$ pc).  Imaging of the Orion nebula (M42, NGC 1976)
was obtained to provide a check on the spectrophotometric reductions
and subsequent interpretations.

\subsection{Observations}

N70 and the Orion nebula were imaged with the CTIO 1.5-m telescope and Rutgers
Fabry-Perot camera between 28 October and 1 November 1993.  The imaging
system consisted of a piezoelectrically controlled scanning etalon and a
servocontroller, both made by Queensgate Instruments, Ltd. (Atherton et al.\
1981), an interference filter with $\approx$ 100\AA\ bandwidth to block out
multiple interference orders, re-imaging optics, and a CCD detector.  Instead
of the high-spectral resolution etalons (FWHM $\approx$ 1--2\AA) that are
standard with the Rutgers Fabry-Perot camera, we installed lower-resolution
etalons (FWHM $\approx$ 5--30\AA) normally resident in the Goddard
Astronomical Fabry-Perot Imaging Camera (GAFPIC, Brown et al.\ 1994).  These
etalons, also made by Queensgate, are optimized for imaging nebular
emission in a variety of lines rather than mapping the nebular kinematics in
just one line.  The calibration of etalon spacing as a function of wavelength
was obtained by scanning the line emission from spectral lamps.  This was done
several times each night, thereby attaining tuning accuracies of less than
$\pm$1\AA.

Initial observations were made with the Goddard ``blue'' etalon and CTIO's
Tek\#4 512$\times$512 CCD chip, where the image scale was 1\farcs1
pixel$^{-1}$ and the total field of view was circular with a diameter of
7\farcm2.  The ``blue'' etalon was tuned to a bandwidth of 7$\pm$1\AA\ (FWHM)
for imaging the \oiii\ emission and green continuum.  During the second half of
the run, observations were made with the Goddard ``red'' etalon and the Tek
1024 \#1 chip; the image scale was 0\farcs96 pixel$^{-1}$ and the resulting
field of view was circular with a diameter just over 7\arcmin.  The Goddard
``red'' etalon was tuned to a bandwidth of 6$\pm$1\AA\ (FWHM) for the \ha,
\nii, and \sii\ observations.  The average radial velocity of N70 with respect
to the local standard of rest is $\sim290$ \kms\ (\cite{dop81}), resulting in a
redshift of the \oiii$\lambda5006.8$ line to 5011.6\AA,
\ha$\lambda6562.8$ to 6569.1\AA, \nii$\lambda6583.6$ to
6590.0\AA, \sii$\lambda6717.0$ to 6723.5\AA, and
\sii$\lambda6731.3$ to 6737.8\AA.  Atmospheric seeing averaged
2\farcs2 (FWHM), enabling spatial resolution on the order of 0.6 pc, assuming
a distance of 55 kpc to the LMC ($m-M=18.7$; \cite{fea97}).  Sky transparency
varied from clear to heavy cirrus.  A summary of the imaging is presented in
Table \ref{tobs}.

\placetable{tobs}

\subsection{Reductions}

Overscan fitting and subtraction, residual bias averaging and
subtraction, domeflat averaging and division, and skyflat smoothing and
division were carried out using the CCDRED routines within
IRAF\footnote{IRAF is distributed by the National Optical Astronomy
Observatories, which is operated by the Association of Universities for
Research in Astronomy, Inc., under contract to the National Science
Foundation.}. The standard and target frames were air-mass corrected
using the mean extinction coefficients at Cerro Tololo (\cite{sto83}).
Cosmic rays were removed with the {\tt cosmicrays} task, which can be
used to identify cosmic rays in fields where only one image has been
obtained.


The \ha, \sii, \nii, and red continuum images were calibrated with
spectrophotometric standards taken close in time to the N70
observations (LTT 9239 and LTT 2415; Hamuy et al.\ 1992, Stone and
Baldwin 1983).  The standards were chosen because they have weak Balmer
absorption lines, so their flux ($F_{\lambda}$) is almost constant in the
wavelength region of interest.  We interpolated between the published
fluxes (in 50\AA\ bins) to arrive at the predicted flux through our
6\AA\ bandpasses.  The standard star calibration error is estimated at
$\lesssim$15\%.

The images of the Orion nebula were taken during cirrus and intermittent
clouds, so calibration with standard stars was not possible.
Therefore, we bootstrapped our Orion calibrations to the wide-field
images at \ha, \nii, and \sii\ (FWHM = 15\AA, 15\AA, 36\AA,
respectively) that were obtained by Walter et al.\ (1992) under
photometric skies; this procedure should yield photometric accuracies
of approximately 10\%.

Because the \oiii\ and green continuum images of N70 were taken in
cirrus, they could not be calibrated with observations of standard
stars.  Instead, an approximate calibration was constructed based on
the observed spectral types of the ionizing stars in N70.  The spectral
energy distributions of six stars classified between O5 and B0
(\cite{oey96a}) were estimated based on the Silva and Cornell (1992)
library of stellar optical spectra.  The closest spectral type in the
catalog was chosen for each star, and the flux at the green continuum
(5031\AA) was predicted based on the calibrated flux at 6751\AA.  The
green continuum image was calibrated to yield the correct stellar
fluxes in the presence of foreground reddening (\ebv=0.06), and the
\oiii-band image was matched to it.  The scatter in measured
6751\AA\ to 5031\AA\ flux is about 20\%, so we can determine the green
continuum calibration only to this accuracy at best.  Regardless of the
uncertainty in the absolute calbration, this \oiii\ image of N70 is the
first to be published, revealing unique morphological characteristics
that have been previously unrecognized.


Once the individual images were calibrated and sky background
subtracted, the closest continuum image was subtracted from the on-line
image to form an emission image.  These emission maps are presented in
the next section.  One of the difficulties with Fabry-Perot imaging is
that the central wavelength of the bandpass changes with radial
distance from the center of the field of view.  The stellar spectra are
essentially flat over the wavelength range of the individual
observations, so the continuum images are unaffected.  However, the
emission-line images are subject to an increasing blueshift with field
angle, amounting to $\sim$4\AA\ at the edge of the field (Atherton et
al.\ 1981; Caulet et al.\ 1992).  The net effect of this shifting of
peak transmissivity  is to decrease the monochromatic sensitivity at
the edge of the field.

To correct for this effect, post-observation ``spectral flat fields''
were made using images of the Orion nebula that were taken with
fixed-wavelength interference filters (Walter et al.\ 1992).  The
spectral flats were created by dividing Walter et al.'s Orion images by
the corresponding Fabry-Perot images and smoothing the results.  This
was done with the continuum-subtracted images in \ha, \nii, and
\sii($\lambda6717+\lambda6731$).  Because Walter et al.'s \sii\ image
includes both \sii\ doublet lines, the individual \sii\ images could
not be corrected separately; however, the individual corrections should
be very similar, so the line ratio $\lambda6717/\lambda6731$ is
basically unaffected.  After correction by these spectral flats, the
total net line fluxes of N70 increased by factors of 1.25 in \ha, 1.05
in \nii, and 2.32 in \sii.  The large increase in \sii\ flux is due to
the concentration of \sii\ emission at the edges of N70, near the edge
of the field.  Line ratios were not as affected, changing by less than
10\% at the edge of the field for \nii/\ha\ and up to 50\% for
\sii/\ha.  No corrections were made to the \oiii\ images.


Reddening corrections were determined based on Oey's spectroscopy and
photometry of stars in LH 114 (the association inside N70); her median
\ebv=0.06 (\cite{oey96a}) was used with reddening coefficients from
Mathis (1990) ($R_V=3.1$) to determine the interstellar reddening and
extinction (0.2 magnitudes in $V$).  When comparing line ratios, it is
worth noting that Dopita et al.\ (1981) used $A_V=0.6$ magnitudes,
which over-corrects in the blue.

Intensity ratio maps, presented in the next section, were created  by
dividing the dereddened emission images.  The line ratios that are
tabulated and discussed are given for specific regions in the nebulae,
where the aperture photometry was done on the emission line images and
then the summed fluxes were ratioed.  Both observed and dereddened
\ha\ fluxes are presented; the low level of reddening derived for N70
means that dereddened \nii/\ha\ and \sii/\ha\ ratios remain essentially
unchanged from the observed values.

\section{Emission-Line Maps and Spectrophotometry \label{maps}}

Observations of N70 over the past twenty years have revealed an ionized shell
with remarkable spectrophotometric properties.  The current Fabry-Perot data
provide observations of the main circular structure of N70 at
\ha, \nii, \sii, \oiii, and their respective line ratios.  The FWHM of stellar
images is 2.3 pixels, or 2\farcs2.  At an LMC distance of 55 kpc, this allows
spatial resolution on the order of 0.6 pc.

\subsection{Maps}

Figure \ref{n70hacont}a is the final dereddened \ha\ image of N70.  The
locations of the two O stars D301-1005 and D301NW-8 (notation of Oey (1996a))
are marked in it and all subsequent emission images and ratio maps.  These
stars are marked with the larger crosses in Figure \ref{n70hacont}b, the
6536\AA\ continuum.  The labeled stars in this figure are those which Oey
(1996a) classified; their spectral types are noted.  Oey chose stars with a
reddening-free index $Q=(U-B)-(B-V)\leq-0.70$ and $V\leq16.0$; her sample
should be reasonably complete in O to early B stars.  The stellar content will
be discussed more fully below.  Note the lack of nebular emission in
the continuum image, indicating that N70 is truly an emission-line nebula with
negligible scattered continuum light.

\placefigure{n70hacont}

The other three N70 emission-line images \nii$\lambda$6584,
\sii$\lambda6717+\lambda$6731, and \oiii$\lambda$5007 are presented in Figures \ref{n70nso}a, \ref{n70nso}b, and \ref{n70nso}c, respectively.  The \sii\ and \oiii\ images are at the original
resolution, but \nii\ was smoothed with a 2 pixel (1\farcs92) Gaussian to
increase the signal-to-noise in the displayed image.

\placefigure{n70nso}

The \ha, \sii, and \oiii\ images were combined to form the color-coded image
shown in Figure \ref{n70color}.  Blue is \oiii, green is \ha, and
red is \sii.  The colors were scaled so that features of all three emission
lines could be seen.  This picture, along with Figures \ref{n70hacont} and
\ref{n70nso}, shows that despite the overall circular symmetry of N70,
significant and unique substructure exists at each emission line.

\placefigure{n70color}

Most notably, strong \ha\ and \oiii\ emission is evident interior to the outer
shell, whereas the \sii\ image shows no such central concentration of
emission.  The interior emission is closely associated with the hottest O-type
stars (labeled in Figure \ref{n70hacont}b), thus indicating a stellar power
source for the nebula sufficient to ionize oxygen twice ($E > 35.1$ eV).  For
reasons more fully explored in Section 4.3, the interior \ha\ and
\oiii\ emission is most likely due to photoionization from the hot stars and
subsequent excitation of the \oiii\ by nebular electrons.

Away from the centrally concentrated hot stars, the nebular emission shows
filamentary substructure in what appears to be a weblike morphology.
Lozinskaya (1992) notes that this sort of emission structure indicates shock
processes at work in the presence of irregularities in the ambient medium.
Many of the well-observed ``mature'' supernova remnants---such as the Veil
Nebula in Cygnus, S147 in Taurus, and the Vela and Gum nebulae
(\cite{loz92})---are characterized by similar filamentary structure in the
same emission lines.

Figure \ref{orionem} shows the central portion of the Orion Nebula in \ha,
\nii, and \sii\ emission.  The Trapezium stars ($\theta^1$ Ori) as well as
$\theta^2$ Ori A and B were saturated in the long \sii\ exposure and so were
not effectively subtracted out of the resulting emission-line images (Figure
\ref{orionem}c).  Only $\theta^1$ Ori C and $\theta^2$ Ori A were saturated in
the \nii\ image, and none were saturated in the short \ha\ exposure (Figure
\ref{orionem}a), probably because of thicker cirrus during that exposure.  To
bootstrap calibrations,  all comparisons between our images and those of
Walter et al.\ (1992) were done in regions away from these bright stars.
Ghost images of the Trapezium stars are barely discernible in the southwest
part of the emission images, and of $\theta^2$A in the northwest.  These are
especially apparent on the \sii\ image.  (Only one of the stars in N70, near
the edge in the north-northwest, was bright enough to cause a noticeable ghost
image.)  The Orion images provide important checks on our reduction methods as
well as spectrophotometric benchmarks for diagnosing the line emission in N70
(discussed more fully in Section 4.1).  Pogge et al.\ (1992)
have presented spectrophotometric Fabry-Perot observations of the Orion Nebula
including ratio maps very similar to those that were constructed from our
data; our data agree well with these published maps and ratios and hence are
not presented herein.

\placefigure{orionem}

Unlike the situation with Orion, intensity ratio maps of N70 have not
been previously available.  Figure \ref{n70ratios} presents three
intensity ratio maps useful for our analysis.  Figure \ref{n70ratios}a
is \nii/\ha, Figure \ref{n70ratios}b is
\sii($\lambda$6717+$\lambda$6731)/\ha, and Figure \ref{n70ratios}c is
\oiii/\ha.  The \sii/\ha\ map is not smoothed in order to make the
change in \sii/\ha\ across the individual filaments clearer.
\nii/\ha\ and \oiii/\ha\ do not show this fine structure, so to
increase signal-to-noise, the maps have been Gaussian smoothed
($\sigma=2$ pixels).

\placefigure{n70ratios}

\subsection{Spectrophotometric Results}
In addition to the intensity ratio maps, emission-line flux ratios were
determined for sections of the nebulae using polygonal aperture photometry.
This method averages over large areas of emission, but allows larger
signal-to-noise measurements and comparisons.  Figures \ref{polygons}a and b
label the polygons used in N70 and Orion, respectively.  The flux in each of
these regions was summed in the individual emission-line images and then
ratioed; the \ha\ fluxes and line ratios with respect to \nii, \sii, and
\oiii\ are presented in Table \ref{n70tab} for N70, and \ha\ fluxes and line
ratios with respect to \nii\ and \sii\ are presented in Table \ref{oriontab}
for Orion.

The uncertainties in the ratios due to the determination of the
background is estimated to be less than 30\% for \nii/\ha, less than
15\% for \sii/\ha\ near the bright rim of N70, and less than 30\% for
\sii/\ha\ in the central region where the \sii\ flux is the lowest.
The ``spectral flat'' correction changed the measured \nii/\ha\ ratios
by less than 10\% regardless of their location in the field of view.
The \sii/\ha\ ratios were decreased by $\sim$ 5\% at the center of the
field by the correction; the correction to the \sii/\ha\ ratio then
increases radially up to 100\% at the N70 rim.  The effects are most
severe for the northern rim.  For example, the \sii/\ha\ ratio of
regions 3, 4, and 5 are increased by 40\% by the ``spectral flat''
correction and region 3 by almost 100\% while region 6 is increase by
only 25\%.  However, the uncertainties in the measured ratios are by no
means this large; the errors are dominated by uncertainty in the value
of the background.

\placefigure{polygons}

\placetable{n70tab}

\placetable{oriontab}

The emission measure for each of the N70 apertures was calculated from
the \ha\ surface brightness, $I_{H\alpha}$.  The emission measure is
defined as $EM = \int n_e^2\,dl$, where $l$ is the column depth of the
emitting material and $n_e$ is its electron density.  For gas with singly
ionized hydrogen and helium,
$n_e \approx 1.1 n_H$. The number of protons $n_H$ can be calculated from
the \ha\ flux.  Assuming $T_e=10^4$ K, $EM$ (\emm) $= 4.4 \times
10^{17} \times I_{H\alpha}$ (erg~cm$^{-2}$~sec$^{-1}$~arcsec$^{-2}$).
With the definition of the emission measure and a simple spherical
shell model, analytic relationships between emission measure and root
mean square (rms) electron density can be derived.  The simplest are
for a line of sight through the center of the shell and for a line of
sight through the edge of the shell.  The first of these relationships,
using emission measures from the center of the optical shell, is

\[ n_{rms}^2 = \frac{EM_{center}}{2\Delta R_s} \]

\noindent where $\Delta R_s$ is the shell thickness.  The second
relationship, using emission measures from the bright rim of the shell,
is

\[ n_{rms}^2 = \frac{EM_{rim}}{2R_s} \left[ 2 \frac{\Delta R_s}{R_s} - \left(
\frac{\Delta R_s}{R_s} \right) ^2 \right] ^{-1/2}  \]

\noindent where $\Delta R_s$ is the shell thickness and $R_s$ is the
radius of the shell, measured out to the edge of the optical emission.

From our N70 observations, we determine a shell radius of
approximately 3\farcm2, or 51 pc, and a shell thickness of 7\farcs5, or
2 pc.  The shell thicknesses were determined from visual examination of
radial plots of the \ha\ emission, where the thickness was set to the
full-width quarter-max of the radial profile of emission across the
filaments.  Using (dereddened) emission measures near the center of
N70, but away from the central two knots, yields $n_{rms} \sim 7.5$~\cc;
emission measures near the rim results in $n_{rms} \sim 4$--6~\cc.
The volume filling factor $f = n_{rms}^2/n_e^2$, where $n_e$ is the
density in the ``clumps'' of gas in the shell and can be determined from
measurements of \sii\ lines as discussed later in this section.

The \ha\ flux measurements can also be used to deduce the Lyman continuum
necessary to ionize the gas in N70.  Assuming the Case B hydrogen recombination coefficient, 

\[ N_{Lyc} = \frac{L_{H\alpha}}{h\nu_{H\alpha}} \frac{\alpha_B}{\alpha_{H\alpha}} = 7.3 \times 10^{11} \times L_{H\alpha} \]

\noindent  for $T = 10^4$ K.  The dereddened \ha\ luminosity of N70 is
$4.3\times10^{37}$ ergs~sec$^{-1}$, so the number of ionizing photons needed
is $3.1\times10^{49}$ s$^{-1}$.  

The luminosity of Lyman continuum photons produced by the stars in the central
cluster can be estimated based on Oey's (1996a) spectral classifications.  The
spectroscopically identified stars marked on Figure \ref{n70hacont}b include
one each of O3If, O5III, O7V, O8III, O9V, and two O9.5V stars as well as
several B stars.  Using $\log T_{eff}$ and $M_{bol}$ from Oey's Table 4, radii
of the stars can be calculated.  (Because we choose to use a distance modulus
of 18.7 instead of 18.4 as Oey did, we subtracted 0.3 from the $M_{bol}$ in
the table.)  Masses of these stars are estimated from her H-R diagram in order
to approximately determine $\log g$.  Using $T_{eff}$, $\log g$, and assuming
$\log \frac{Z}{Z_{\odot}} = -0.3$, the Kurucz (1992) model atmospheres can be
used to estimate the flux of ionizing photons (cm$^{-2}$ s$^{-1}$) from the
stars.  The Lyman continuum luminosity for each star is then calculated using
the radius determined above.  The total ionizing luminosity from the cluster
is $7.0 \times 10^{49}$ photons s$^{-1}$, twice as much as is necessary to
ionize the interstellar hydrogen in N70.  The ultimate fates of any EUV
photons beyond what is necessary to ionize the emission nebula can include
absorption by dust, absorption by other atomic species, and escape from the
nebula.  As expected, the late O and early B stars contribute very little to
the ionizing radiation; the O3If and O5III stars alone contribute over 60\% of
the total Lyman continuum.  Varying the choice of stellar metallicity between
$\log \frac{Z}{Z_{\odot}} = -1.0$ and 0.0 changes the ionizing flux by less
than 5\%.

Another way to estimate the Lyman continuum luminosity from the cluster
is with the compilations of Vacca et al.\ (1996).  Stellar parameters
such as radius, mass, absolute magnitude, radius, and Lyman continuum
flux (both photons cm$^{-2}$ s$^{-1}$ and total  s$^{-1}$) are
tabulated by spectral type and luminosity class.  Based on the spectral
classification of the N70 stars, the ionizing luminosity determined
from Vacca et al.\ (1996) is $2.2\times10^{50}$ photons s$^{-1}$, seven
times higher than the hydrogen ionization rate in N70.  These
luminosities are for solar-metallicity stars, but as determined above,
the metallicity only slightly affects the amount of Lyman continuum
radiated by the stars.

The factor of three difference from the previous determination is because Vacca
et al.\ calculate radii almost twice as large as we found.  The root of this
difference is in the bolometric magnitudes assumed for each spectral type.
Oey (1996b) determined bolometric corrections in the manner described by
Massey et al.\ (1995) using broadband colors.  Vacca et al.\ (1996) discuss
their calibration of $M_V$ with spectral type; in general they determine {\em
brighter} absolute magnitudes.  For the two hottest stars, the differences are
0.9 and 1.9 magnitudes, which translates into more luminous and thus larger
stars for the same effective temperature.  The discrepancy between the two
determinations of bolometric magnitude and thus ionizing luminosity shows that
this can be a tricky business, and should illustrate the large uncertainties
in the calculated ionizing luminosity.  However, our predictions of ionizing
luminosity are quite consistent with that recently tabulated by Oey and
Kennicutt (1997).

As discussed at length in Section \ref{or_neb}, the line ratios in
Table \ref{n70tab} and Figure \ref{n70ratios} are discordant with a
straightforward photoionization model.  The \nii/\ha\ ratios in Table
\ref{n70tab} are similar to those in the central regions of Orion, but the
\sii/\ha\ ratios are much higher than expected for a photoionized nebula.

In addition, there is a definite increasing trend in the \sii/\ha\ intensity
ratio across the individual filaments on the face of N70 in addition to the
overall increasing \sii/\ha\ intensity ratio with increasing radius from the
center of the nebula.  This can be seen in Figures \ref{n70nso}b and
\ref{n70ratios}b.  Figure \ref{filaments} displays cuts across three filaments
in N70; the \sii/\ha\ ratio rises with distance from the center of N70.  A
fourth plot shows a cut across the Orion Bar; the \sii/\ha\ ratio is
multiplied by a factor of ten in order to see it on this scale.  As discussed
below, the Bar is an ionization front in the Orion nebula seen edge on.  The
cut across the bar has similar shape as cuts across N70 filaments, but the
\sii/\ha\ ratios are much smaller.  Also the physical scale is {\em much}
smaller in Orion; 16 pixels total only 0.03 pc while 16 pixels totals 4.1 pc
linear distance in N70.

\placefigure{filaments}

The discussion of the reduction of the N70 \sii\ images focussed on the sum of
the individual \sii$\lambda$6717 and \sii$\lambda$6731
images.  Although the ``spectral flat fields'' could not be individually derived
for these two line images, we can assume that the radially decreasing ``gain''
in each of these images is similar and thus measure the density-dependent
\sii$\lambda$6717/$\lambda$6731 line ratio (\cite{ost89}).  As
Table \ref{n70tab} shows, these ratios range from 1.4 to 1.9; for $T=10^4$ K,
the maximum $\lambda$6717/$\lambda$6731 line ratio in models
is about 1.4 at the low density limit.  Therefore, the density of gas in N70
is low, probably $<$ 100~\cc, in {\em all} parts of the optical nebula.
Because we can not constrain $n_e$ any further, we constrain the
filling factor in the shell only to be $>$ 0.002--0.006 using the rms
densities determined from the emission measure of the gas.  If $n_e$
were only 10~\cc, the filling factor in the shell would be $\sim$ 0.2--0.6.

\section{Comparison with Orion, LMC Emission-Line Regions, and Nebular Models \label{or_neb}}
\subsection{Orion \label{orion}}

When comparing N70 to the Orion Nebula, it is important to remember how
much more detail we can see in nearby objects---Orion is $\sim450$ pc
away (\cite{gou82}), 120 times closer than N70.  At the distance of the
LMC, the part of the Orion Nebula seen in our images would be contained
in 3.5 pixels!  In addition, our coarse resolution (pixel scale
0\farcs96/pixel and seeing $\sim$2\farcs2) limits the amount of fine
structure we would be able to see in our N70 images, especially when
compared with Orion and LMC emission regions which have been observed
with WFPC2.

As previously mentioned, Pogge et al.\ (1992) present two-dimensional
images of the central region of Orion.  The complexity of these images
is apparent; the central bright regions are certainly ionized by the
Trapezium stars, but there are also Herbig-Haro objects, the ``Dark
Bay'' (very high absorption), and the ``Bar''.  Wen and O'Dell (1995)
have constructed a three-dimensional model of this region of the Orion
Nebula.  In their model, Orion is a ``blister'' on the edge of the
giant molecular cloud OMC-1 rather than a classical Str\"omgren
sphere.  Wen and O'Dell (1995) find that most of the radiation from the
Orion Nebula arises from a relatively thin surface layer of ionized
material, and have used that information to model the geometry near the
Trapezium ($\theta^1$ Ori).  They show that the emission enhancement
seen just to the west of the Trapezium is due to the ``hilly''
structure of the ionized layer; the bright emission comes from an area
which is closer to the ionizing stars than average.  The Dark Bay is an
area where the ``lid'', or the ionization front between an Earth
observer and the Trapezium, is thicker than average and has increased
absorption by neutral gas and dust.  The Bar, on the other hand, is the
ionization front seen edge on.  This is consistent with the thinness of
the Bar in \nii\ and \sii\ compared to \ha\ (see Figure \ref{orionem})
and with the enhancements in the \nii/\ha\ and \sii/\ha\ that would
arise in such a transitional layer where sulfur and nitrogen have yet
to reach their highest ionization stages (c.f.  \cite{pet95}).  The
\oiii\ image of Pogge et al.\ (1992) shows that the higher-ionization
material is mainly on the northwest side of the Bar, closer to the
ionizing stars.  The \nii - and \sii -bright objects which are to the
southeast of the Bar near $\theta^2$ Ori A are the shock-excited
Herbig-Haro objects M42-HH3 (closer to the Bar) and M42-HH4 (a little
further to the southeast).

Figure \ref{orionn70} is a plot of the emission line ratios for Orion and N70
from Tables 2 and 3.  The differences between the two emission regions are
obvious:  the Orion Nebula, even in the shock-excited Herbig-Haro objects, has
a moderate range of \nii/\ha\ and very low \sii/\ha\ throughout, while N70 has
consistently {\it lower} \nii/\ha\ by $\sim 50$\% but a wide range of
\sii/\ha\ ratios.  As seen in Table 3 and Figure \ref{polygons}b, the
\nii/\ha\ ratios across the face of Orion are lower than those in the Bar or
in the Herbig-Haro objects.  The regions of N70 with low \sii/\ha\ have
\nii/\ha\ most similar to that seen in Orion; Table 2 and Figure
\ref{polygons}a show that these regions are close to the ionizing stars and
hence are most likely photoionized like the central regions of Orion.  This is
corroborated by Figure \ref{n70nso}c, which shows that \oiii, a
higher-ionization state, is quite elevated near the central stars that power
N70.  There is disagreement over whether or not Orion has lower metallicity
than solar (\cite{wal92}), but in either case the average LMC nitrogen
abundance of 0.4 solar (\cite{rus92}) is significantly less than that of
Orion, which can explain the lower \nii/\ha\ ratios seen in N70.

\placefigure{orionn70}

The enhanced \sii/\ha\ emission-line ratios across Orion's Bar and the
outer filaments in N70 lead one to wonder whether the filaments in N70
are actually just ionization fronts like the Orion Bar.  However, as
mentioned above, the physical scale and relative amount of
\sii\ emission is very different in N70 than in Orion.  An ionization
front in N70 would occur over physical lengths too small to be
distinguished due to the effects of seeing.  Another difference is
apparent when comparing Figures \ref{n70hacont}a and \ref{n70nso}b with
Figures \ref{orionem}a and \ref{orionem}c: while the width of the
filaments in N70 is approximately the same whether measured in \ha\ or
\sii, the same is not true for the Orion Bar.  The Bar's \ha\ emission
extends over a much larger area than its \sii\ emission, consistent
with the \sii\ tracing the transitional ionization front.   The similar
\ha\ and \sii\ widths in the filaments of N70 indicate that the
elevated \sii/\ha\ ratios are not due to multiple ionization fronts
within the nebula.  Instead, {\em the displacement in \ha\ and
\sii/\ha\ peaks observed in N70 is consistent with an outward
propagating shock front that is ``back illuminated'' by the embedded
cluster of OB stars} (see Figures \ref{n70ratios}b and
\ref{filaments}).  The back illumination would be responsible for the
inward-facing \ha\ peak, while the shock front would explain the
enhanced \sii\ emission.

The special ionization and excitation mechanisms required for the elevation of
the \sii/\ha\ intensity ratios can be constrained by comparison with other
emission nebulae and with theoretical models.

\subsection{LMC Emission-Line Regions}

When comparing N70 with objects in the Large Magellanic Cloud, differences in
metallicities can be assumed to be small.  Observed emission-line ratios of
various objects in the LMC were collected from the literature and plotted in
Figure \ref{shells}.  The supernova remnants are from Danziger and Leibowitz
(1985) with the exception of two from Westerlund and Mathewson (1966).  The
\hii\ region line ratios are those observed by Wilcots (1992).  Hunter (1994)
conducted a survey of ionized shells and supershells in the LMC; her ``giant
shells'' have radii 50 to 300 pc, and her ``supergiant shells'' have radii
greater than 300 pc.  The data plotted are for a variety of positions within
three giant shells and three supergiant shells rather than the average line
ratio across these shells.  The N70 ratios in Table \ref{n70tab} are also
plotted.   As shown in  Figure \ref{shells}, the \nii/\ha\ and \sii/\ha\ flux
ratios found in N70 span a range of values similar to that of giant and
supergiant shells.  However, a significant population of regions show even
higher \sii/\ha\ flux ratios that are more characteristic of supernova
remnants (SNRs).

\placefigure{shells}

Unlike the comparison with the Orion Nebula, the line ratios in the central
regions of N70 are very similar to those of \hii\ regions in the LMC.  The
\hii\ regions in Wilcots' (1992) sample are ``classical'' \hii\ regions with
``interesting'' morphologies.  Their \ha\ luminosities are in the range of
$10^{36}$ to $10^{37}$ erg s$^{-1}$, a bit lower than N70.  However, the
\ha\ luminosity of the two central knots (1 and 2 in Figure \ref{polygons}a)
is $5.3 \times 10^{36}$ erg s$^{-1}$, similar to Wilcots' \hii\ regions.  The
line ratios of these regions of N70 are also similar to some of Hunter's
(1994) giant shells; she identifies these as \hii\ regions within the shells.
We believe the same is true for the region of N70 directly encircling the
ionizing stars; these are volumes where photoionization is the main influence
on the gas.  Of course the massive stars are blowing winds into the gas, but
there is nothing to distinguish these regions from other nebulae interacting
with massive main-sequence stars.

The line ratios of N70 share some properties with the sample of SNRs; however,
many of the SNRs have higher \nii/\ha\ but similar \sii/\ha.  As discussed
below, these regions perhaps have higher shock velocities than those important
in N70.  This comparison shows that the very high \sii/\ha\ ratios measured in
N70 are not unknown in LMC emission regions.  The exception is the point at
\sii/\ha=2.06, which is the north rim of the nebula, and an area where the
shape of the rim is not circular.  It is possible that this is a volume
expanding into lower density material, or that there is some off-center effect
of winds or a supernova.

The more intriguing comparison is with the giant and supergiant shells
(\cite{hun94}).  Like N70, these shells show a range of both \nii/\ha\ and
\sii/\ha\ ratios.  The trends in these line ratios are very similar for all of
the shells; some regions of the shells are more like \hii\ regions, and others
are more like SNRs.  In her paper, Hunter presents many plots of diagnostic
line ratios.  Although the data presented here does not include many of the line
ratios that Hunter examined, this spectrophotometric information can be collected
from the literature on N70 (\cite{las77}, \cite{las81}, \cite{dop81}) and
compared with the giant and supergiant shells.  For example, N70 line ratios
such as \oiii/\hb\ and \oii/\oiii\ are also consistent with the values
measured for Hunter's sample.

\subsection{Nebular Models \label{neb}}

A complete modeling of the photoionization and shock 
conditions in N70 is beyond the scope of this paper.  However, comparison with
existing models can yield significant insights on the respective roles played
by stellar photons and shock waves.
 
Shull and McKee (1979) constructed theoretical models of interstellar
shocks moving through a low-density medium.  Their models pre-ionize
the gas in front of the shock with UV flux created by the shocked gas
itself.  In slow shocks, such as those to be considered below, the gas
flowing toward the shock is only partially pre-ionized.  The strength
of the Balmer lines are sensitive to the pre-ionization, but metal-line
strengths depend more strongly on collisional excitation by electrons
behind the shock.  With solar abundances, pre-shock density $n_0
=10$~cm$^{-3}$, and shock velocity of 40 \kms, the Shull and McKee
(1979) models predict \nii/\ha\ = 0.02 and \sii/\ha\ = 1.24.  One of
the models also explores the effect of depleted abundances on line
ratios, predicting that \oi, \Ni, and \sii\ are strengthened while
\oii, \oiii, and \nii\ are weakened relative to \ha\ due to diminished
cooling and a larger hydrogen recombination zone.

Hartigan et al.\ (1994) examine slower shocks, down to 15 \kms, but with
higher pre-shock densities ($n_0=10^2$, $10^3$, and $10^4$ cm$^{-3}$).  These
models show the \sii/\ha\ flux ratio 
increasing with decreasing shock velocity, reaching  
a peak value of $\sim$2 at a shock velocity of 25 \kms in the 
10$^2$ cm$^{-3}$ model.  In
addition,  \nii/\ha\ decreases and \Ni/\ha\ increases with decreasing shock
velocity, as would be expected if the shock no longer has the energy to ionize
the nitrogen.  (S$^0 \rightarrow$ S$^+$ requires only 10.4 eV while N$^0
\rightarrow$ N$^+$ requires 14.5 eV).  Extending these results to the lower
pre-shock densities estimated for N70 (0.1 -- 0.7 cm$^{-2}$; Rosado et al.
1981; Meaburn 1978) would suggest further enhancement of \sii\, as collisional
de-excitation can be ignored. 

The high \sii/\ha\ ratios seen in N70 can be explained with the
aforementioned models with shock velocities of 25 to 40 \kms,
consistent with the reported expansion velocities of 20 -- 40
\kms\ (\cite{las77}, \cite{bla80}) but less than the expansion velocity
of 70 \kms\ found by Rosado et al.\ (1981).\footnote{\cite{dop81}
argues that the velocity field in N70 cannot be interpreted as simple
expansion.  Even without coherent expansion, the observed velocity
dispersion is sufficient to provide collisional (shock) excitation
consistent with the models.}  The low \nii/\ha\ ratios in the outer
parts of N70 are also consistent with this interpretation.  However,
the \oiii$\lambda5007$/\hb\ and
\oii$\lambda\lambda$3726,3729/\hb\ measured by Lasker (1981) and Dopita
(1981) as well as the \oiii/\ha\ presented here are higher than the
flux ratios predicted by the low-velocity shock models.  We therefore
propose that N70 has distinct regions of photoionization augmented to
varying degree by shock ionization and excitation.  This composite
powering is qualitatively apparent by comparing the morphology of the
\sii\ emission with the \ha, \nii, and \oiii\ emission morphologies,
which are more similar to each other than to the shock-excited \sii.

The unusually bright \oiii\ emission seen in the southern edge of N70
cannot be explained by the proposed photoionization plus slow-shock
model. 
Perhaps a stronger shock has excited the gas on the southern rim. X-ray data
from the Einstein satellite 
(Chu and Mac Low 1990) are suggestive of an off-center supernova
within the bubble, as their model explains.  The detection is marginal,
but the enhanced \oiii\ emission could come from a higher velocity
shock than that exciting the \sii\ on the western and northern rims of
N70.

\subsection{Radiative and Mechanical Energetics}

In her study of ionized bubbles in the LMC, Oey (1996b) found that those with
high X-ray luminosities and large expansion velocities have discrepantly small
radii relative to their expansion ages and stellar wind powering.  She
concluded that the dynamical discrepancy is probably caused by recent
supernova events accelerating the shells to higher expansion velocities than
would be obtained by stellar winds alone.  She also suggested that additional
energy sink mechanisms might explain the anomalously smaller radii.  Because her
sample of ``superbubbles'' included N70, further constraints on the relevant
energetics can be obtained by considering both the mechanical and radiative
luminosities that are involved.

\subsubsection{Sources}

As shown in Figure 1b, N70 contains 17 massive stars with spectral types
ranging from B2.5V to O3If (Oey 1996a).  The total wind luminosity from this
population was modeled by Oey (1996b) to be no more than 10$^{37}$ erg
s$^{-1}$ (see her Figure 1).  A total mechanical luminosity of 1.5 $\times$
10$^{37}$ erg s$^{-1}$ is obtained using the spectral classifications in Oey
(1996a), the corresponding luminosities (for Galactic stars) listed in
Leitherer (1997), and adjustments for the LMC's lower metallicity.  Here, we
used the theoretical prediction that $\dot{M} \propto Z^{0.8}$ and $v_\infty
\propto Z^{0.13}$, where $L_w = \dot{M} {v_\infty}^2 / 2$ (Leitherer 1997).
Given this source of power, is it sufficient to explain the currently observed
X-ray luminosity, enhanced \sii\ emission, and expanding motions?

\subsubsection{Sinks}

N70 is noted for having a high X-ray luminosity relative to predictions based
on its size and expansion velocity (\cite{oey96b}, \cite{chu90}).
Observations by the Einstein observatory yield $L_x(Einstein) = 1.8 \times
10^{35}$ erg s$^{-1}$ (\cite{chu90}), which when scaled up by 3 to the
0.1--2.4 keV ROSAT bandpass becomes $L_x(ROSAT) = 5.4 \times 10^{35}$ erg
s$^{-1}$, or about 7 times higher than is predicted from the nebular dynamics
(\cite{chu95}).  For our purposes, it is worth noting that the total X-ray
luminosity comprises a negligible fraction of the total mechanical power that
is available from the stellar winds.

Another, more important, radiative sink of input mechanical power is
the excess \sii\ emission that we measure.  From Table 2, the ratio of
summed \sii\ and \ha\ fluxes is 0.77, with a total \sii\ flux of 9.24
$\times$ 10$^{-11}$ erg s$^{-1}$ cm$^{-2}$, yielding a total
\sii\ luminosity of $3.37 \times 10^{37}$ erg s$^{-1}$.  If
photoionization typically produces flux ratios of $f($\sii$)/f(H\alpha)
\leq 0.4$ (see Figure 9), then $\geq$50\% of the \sii\ emission must
result from other, more mechanical, ionization/excitation processes.
The required powering of $L($\sii$) \ge 1.7 \times 10^{37}$ erg
s$^{-1}$ would be multiplied by about 1.5, if the excess cooling by
\oi\ is included (\cite{dop81}).  These power requirements are
marginally higher than those provided by the stellar winds, leaving
little ``wiggle room'' for any other energy sinks such as nebular
expansion.

A variety of techniques can be used to estimate the kinetic energies and
mechanical luminosities associated with expanding shells of gas (cf.\
\cite{ten88}, \cite{loz92}).  All of these techniques are critically sensitive
to the density and structure of the surrounding medium as well as the powering
timeline, and hence are fraught with uncertainties.  The disparities in size,
age, and expansion velocity found by Oey (1996b) underscore these
difficulties.  Nevertheless, the expansion energetics can be significant and
hence are worth estimating.

In their kinematic study Rosado et al.\ (1981) obtain a swept up mass
of $2.3 \times 10^3 M_{\odot}$ and an expansion velocity of 70 \kms,
thus deriving a kinetic energy of $1.1 \times 10^{50}$ erg.  A more
representative expansion velocity is about 35 \kms\ (\cite{chu88}), resulting in
kinetic energy of $2.8 \times 10^{49}$ erg.  Averaging this energy over
the 5 My lifetime of the cluster would then yield a mechanical
luminosity of $2 \times 10^{35}$ erg s$^{-1}$---comfortably less than
the $10^{37}$ erg s$^{-1}$ available from the winds.

Slightly higher estimates of mechanical energy (E$_m$ = [1 -- 7.5] $\times$
10$^{50}$ erg) are obtained with a momentum-conserving model for the
expansion (\cite{ten88}), where

\[ E_m = 5.3 \times 10^{43} n_0^{1.12} R^{3.12} v^{1.4}, \]

\noindent the expansion velocity is $v \approx 35$ \kms\, and the
ambient density is assumed to be $n_0 \approx 0.1$--$0.5$ \cc.

We conclude that the radiative and mechanical sinks of energy
collectively exceed the input wind power by factors of $\sim 2$, the
observed radiative sink of \sii\ alone being dominant.  Allowing for
other radiative sinks such as \oi, \oii, and \oiii\ in the optical
(\cite{dop81}) and by \cii, \ciibr, \ciii, and \ciiibr\ in the UV (cf.
\cite{dop84}) would further exacerbate the observed disparity in
energetics.  One or two recent supernovae with individual energies of
10$^{51}$ ergs would be sufficient to make up the difference.  Recent
supernova activity would also help to explain the anomalously high
expansion velocity and X-ray luminosity.

\section{Summary \label{summary}}

N70 is a fascinating emission-line region in the Large Magellanic Cloud
whose spherical symmetry belies its complex powering.  The data
presented here cannot solve the mystery of N70's dynamic history, but
can provide new insights on the nebular energetics based on diagnostic
emission line ratios such as \nii/\ha\ and \sii/\ha.  Our conclusions
are as follows:

\begin{itemize}

\item{Although N70's dynamics cannot be well explained by a standard
pressure-driven bubble model (\cite{oey96b}; note that the high luminosity
half of her sample of bubbles are inconsistent with the model), its
emission-line ratios---\nii/\ha\ and \sii/\ha\ from our data and
\oiii/\hb\ and \oii/\oiii\ from the literature---match well with the
ratios of other LMC giant and supergiant shells in the LMC
(\cite{hun94}).\footnote{The samples of Oey (1996b) and Hunter (1994)
do not overlap; a useful endeavor would be to collect emission-line
diagnostics for Oey's sample for comparison with Hunter's data, as well
as to investigate the dynamics of Hunter's sample using Oey's model.} }

\item{N70's central regions emit emission lines with flux ratios similar to
those of photoionized \hii\ regions, while the rim of the N70 shows elevated
\sii\ emission.  The ionization of all of the hydrogen can be attributed to
stellar EUV photons, but additional processes such as slow shocks are necessary to
explain the combination of high \sii\ emission and low \nii\ emission,
especially in the northeast and southern parts of the nebula.}

\item{The energetics associated with the stellar winds, expanding shell,
and radiating \sii\ are best reconciled if one or two supernova explosions 
have occurred within N70 in the past $\approx 10^6$ years.  The enhanced \oiii\ emission and marginal X-ray detection to the south also indicate higher velocity shocks from recent supernovae.}

\end{itemize}

A wealth of information about the small-scale details of ionization and
shock fronts has been gained about other emission regions with HST and
WFPC2 (e.g.\ Hester et al.\ 1995, 1996); some of the remaining
questions about N70 could be answered with higher resolution images,
especially a finer-scale mapping of line ratios across its filaments.

\acknowledgments

We would like to thank Don Walter for sharing his calibrated images of
the Orion nebula.  
We also thank the referee, You-Hua Chu, for her guidance and patience as we
wrestled with the nebular energetics.
BPS would like to thank Paul Hodge, Gene Magnier,
and Eliot Malumuth for helpful advice and conversations while reducing
and analyzing this data.  WHW is grateful to NASA for funding under the
Astrophysics Data Program (\#071-96adp), to the UIT Science Team led by
Ted Stecher, and to the Goddard SNR and ISM lunch bunch led by Robin
Shelton and Jonathan Keohane for intellectual stimulation and support.

\clearpage
\figcaption{(a) Dereddened \ha\ emission-line image of N70 taken with a
Fabry-Perot imaging system with FWHM$\sim$6\AA.  (b) 6536\AA\ continuum; stars
are labeled with spectral types from Oey (1996a).  The field of view has a
diameter of about 7\arcmin, or 110 pc at the distance of the LMC.
\label{n70hacont}}

\figcaption{Dereddened emission-line images of N70: (a) \nii\ (smoothed with a 
2 pixel Gaussian), (b) \sii, and (c) \oiii.  The FWHM is $\sim$6\AA\ for 
\nii\ and \sii, and $\sim$7\AA\ for \oiii.  The field of view has a
diameter of about 7\arcmin. \label{n70nso}}

\figcaption{Color-coded image of N70 constructed from emission-line images.
Blue is \oiii, green is \ha, and red is \sii.  Note that yellow = green + red.
\label{n70color}}

\figcaption{Emission-line images of the central region of Orion, approximately
centered on the Trapezium:  (a) \ha, (b) \nii, (c) \sii.  The field of view 
has a diameter of about 7\arcmin, or less than 1 pc.\label{orionem}}

\figcaption{Emission-line ratio maps of N70: (a) \nii/\ha, (b) \sii/\ha, and
(c) \oiii/\ha.  The \nii, \oiii, and \ha\ maps were smoothed before constructing the \nii/\ha\ and \oiii/\ha\ ratios.  The \oiii/\ha\ map does not account for the radially decreasing sensitivity at \oiii\ (see text).  \label{n70ratios}}

\figcaption{\ha\ images of N70 and Orion showing the polygonal apertures used for aperture photometry of various regions of emission.  Fluxes and ratios are listed in Tables \ref{n70tab} and \ref{oriontab}.  \label{polygons}}

\figcaption{\sii/\ha\ intensity ratios across three filaments in N70, one in the
southeast, one in the south, and another in the west.  These filaments are
characteristic of many of the filaments in N70.  The \sii/\ha\ ratio across
the Orion Bar is also plotted (multiplied by a factor of ten so that it could
be seen more easily).  Each of the four cuts begins on the side of the
filament closer to the ionizing stars (LH 114 for N70 and the Trapezium for
Orion) and extends 16 pixels (15\farcs4) radially away from the center.
15\farcs4 corresponds to 4.1 pc at the distance of the LMC and 0.034 pc at the
distance of Orion.  \label{filaments}}

\figcaption{\nii/\ha\ vs.\ \sii/\ha\ for regions in N70 and the Orion Nebula.  The ratios plotted are those in Tables 2 and 3 from the polygonal apertures
shown in Figure \ref{polygons}.  \label{orionn70}}

\figcaption{\nii/\ha\ vs.\ \sii/\ha\ for a variety of line-emission regions in
the Large Magellanic Cloud.  The line ratios of N70 span the excitation domain
populated by giant shells, supergiant shells, and even SNRs. \label{shells}}

\begin{table}
\dummytable\label{tobs}
\dummytable\label{n70tab}
\dummytable\label{oriontab}
\end{table}

\end{document}